\documentclass{aa}

\usepackage{psfig}
\usepackage{graphicx}
\usepackage{natbib}
\usepackage[english]{babel}  
\usepackage{times}

\begin{document}

\title{X-rays from protostellar jets: emission from continuous flows}

\author{R. Bonito\inst{1, 2} \and S. Orlando\inst{2} \and G. Peres\inst{1}
\and F. Favata\inst{3} \and R. Rosner\inst{4, 5}}

\offprints{R. Bonito\\ \email{sbonito@astropa.unipa.it}}

\institute{Dip. Scienze Fisiche ed Astronomiche, Sez. Astronomia,
Universit\`a di Palermo, P.zza del Parlamento 1, 90134
Palermo, Italy
\and 
INAF -- Osservatorio Astronomico di Palermo, P.zza del Parlamento 1,
90134 Palermo, Italy 
\and
Astrophysics Div. -- Research and Science Support Dept. of
ESA, ESTEC, Postbus 299, 2200 AG Noordwijk, The Netherlands
\and
Dept. of Astronomy and Astrophysics, Univ. of Chicago, Chicago, IL
60637 
\and
Center for Astrophysical Thermonuclear Flashes, University of Chicago,
5640 S. Ellis Avenue, Chicago, IL 60637, USA } 

\date{Received, accepted}

\authorrunning{}
\titlerunning{}

\abstract
{Recently X-ray emission from protostellar jets has been detected with
  both XMM-\emph{Newton} and \emph{Chandra} satellites, but the physical
  mechanism which can give rise to this emission is still unclear.}
{We performed an extensive exploration of a wide space of the main
  parameters influencing the jet/ambient interaction. Aims include: 1) to constrain the jet/ambient interaction regimes leading to the X-ray
  emission observed in Herbig-Haro objects in terms of the emission by a shock forming at the interaction front between a continuous supersonic jet and the surrounding medium; 2) to derive detailed
  predictions to be compared with optical and X-ray observations of
  protostellar jets; 3) to get insight
  into the protostellar jet's physical conditions.}
{We performed a set of bidimensional hydrodynamic numerical simulations, in cylindrical coordinates, modeling supersonic jets ramming a uniform ambient medium. The model takes into account the most relevant physical effects, namely the thermal conduction and the radiative losses.}
{Our model explains the observed X-ray emission from protostellar jets
in a natural
  way. In particular we find that the case of a protostellar jet less
  dense than the ambient medium reproduces well the observations of the nearest Herbig-Haro object,
  HH\,154, and allows us to make detailed predictions of a possible
  X-ray source proper motion ($v_{\rm sh} \approx500$ km s$^{-1}$),
  detectable with \emph{Chandra}. 
  Furthermore our results suggest that the simulated protostellar jets which best reproduce the X-rays observations cannot drive molecular outflows.}
{}


\maketitle

\section{Introduction}
\label{Introduction}

The early stages of the star birth are characterized by a variety of
mass ejection phenomena including collimated jets. These plasma jets
can travel through the interstellar medium at supersonic speed with
shock fronts forming at the interaction front between the jet and the
unperturbed ambient medium.  In the last 50 years these features have
been studied in detail in the radio, infrared, optical and UV bands
and are known as Herbig-Haro (hereafter HH) objects (\citealt{her50};
\citealt{har52}; see also \citealt{rb01}).

\citet{pfg01} predicted that the most energetic HH objects could be
sources of strong X-ray emission.  Following \citet{zr66} one can derive
useful relations between the physical parameters of interest (as the
plasma temperature and the shock velocity) of the post-shock region,
in particular

\begin{equation}
T_{\rm psh} = \frac{\gamma-1}{(\gamma+1)^2} 
\left(\frac{m v_{\rm sh}^{2}}{k_{\rm B}}\right)
\label{eq:rappT}
\end{equation}

\noindent
where $T_{\rm psh}$ is the post-shock temperature, $\gamma$ is the
ratio of specific heats, $v_{\rm sh}$ is the shock front speed, $m$
is the mean particle mass and $k_{\rm B}$ is the Boltzmann constant.
Assuming a typical velocity, $v_{\rm sh}\approx500$ km s$^{-1}$ (as
measured in HH\,154, see \citealt{fld05}), the expected post-shock
temperature is of few millions degrees, thus leading to X-ray emission.

Recently, X-ray emission from HH objects has been detected with both
the XMM-\emph{Newton} and \emph{Chandra} satellites: the low mass young stellar objects (YSO) HH\,$2$ in Orion
(\citealt{pfg01}) and HH\,154 in Taurus (\citealt{ffm02}; \citealt{bfr03}), the high mass YSO objects
HH\,$80/81$ in Sagittarius (\citealt{ptm04}) and HH\,$168$
in Cepheus A (\citealt{pt05}), and HH $210$ in Orion (\citealt{gfg06}).
Indications of X-ray emission from protostellar jets are also
discussed by \citet{tkk04} and \citet{gsb05}.
A summary of the relevant physical quantities observed
for these objects is presented in Tab.~\ref{tab:obs}.

\begin{table*}[!t]
\caption{Relevant physical quantities observed in confirmed X-ray
emitting HH objects, where $L_{\rm X} $ is the reported X-ray luminosity,
$kT$ and $N_{\rm H}$ are the best fit parameters derived from spectral
analysis for the temperature (in keV) and for the interstellar absorption
column density, $v_{\rm sh}$ is the shock front velocity derived from
optical observations, $D$ is the distance of the object observed and $L_{\rm bol}/L_{\odot}$ is the bolometric luminosity of the jet driving source
(\citealt{pfg01}; \citealt{ffm02}; \citealt{bfr03}; \citealt{ptm04};
\citealt{pt05}; \citealt{gfg06}).}\label{tab:obs}
\par
\begin{tabular}{lccccccc}
\hline
\hline
object &$L_{\rm X} $   & $kT$ & $N_{\rm H}$ & $v_{\rm sh}$ & $D$ & $L_{\rm bol}/L_{\odot}$ & $L_{\rm X}/L_{\rm bol}$ \\
       & [$10^{29}$ erg s$^{-1}$] & [keV] & [$10^{22}$ cm$^{-2}$] & [km
s$^{-1}$] & [pc]  & & \\
\hline
HH 2     & $5.2$ & $0.23$        & $\leq 0.09$ & $230$ & $480$  & $81$$^{a}$               & $1.7\times10^{-6}$ \\
HH\,154  & $3.0$ & $0.34$        & $1.40$      & $500$ & $140$  & $40$$^{b}$               & $9.7\times10^{-7}$ \\
HH 80/81 & $450$ & $0.13$        & $0.44$      & $700$ & $1700$ & $2\times10^{4}$$^{c}$    & $1.5\times10^{-4}$ \\
HH 168   & $1.1$ & $0.5$         & $0.40$      & $500$ & $730$  & $2.5\times10^{4}$$^{c}$  & $3.6\times10^{-7}$ \\
HH 210   & $10$  & $0.07 - 0.33$ & $0.80$      & $130$ & $450$  & $-$                      & $-$\\
\hline
\hline
\noalign{\smallskip}
\multicolumn{4}{l}{$^a$ \citet{cwk01}} \\
\multicolumn{4}{l}{$^b$ \citet{lfl05}} \\
\multicolumn{4}{l}{$^c$ \citet{chp06}} \\
\end{tabular}
\end{table*}

In addition to the intrinsic interest in their physics, understanding
the X-ray emission from protostellar jets is important in the context
of the physics of stars and planets formation. 
X-rays (and more in general ionizing radiation) affect many aspects of the environment of young stellar objects and, in particular, the physics and chemistry of the accretion disk and its planet-forming environment. The ionization state of
the accretion disk around young stellar objects will determine its
coupling to the ambient and protostellar magnetic field, and thus,
for example, influence its turbulent transport. In turn, this will affect the
accretion rate and the formation of structures in the disk and, therefore,
the formation of planets. Also, X-rays can act as catalysts of chemical
reactions in the disk's ice and dust grains, significantly affecting
its chemistry and mineralogy.

The ability of the forming star to ionize its environment will therefore significantly affect the outcome of the process, independently from the origin of the ionizing radiation. While all young stellar objects are strong X-ray sources, they will irradiate the disk from the central hole, so that stellar X-rays will illuminate the disk with grazing incidence, concentrating their effects in the central region of the disk (although flared disks
can alleviate the problem to some extent). Protostellar jets are located above the disk, so that they will illuminate the disk with near normal incidence, maximizing their effects even in the outer disk regions shielded from the stellar X-rays. For example, the emission from HH\,154 is located at some 150 AU from the protostar, ensuring illumination of the disk with favorable geometry out to few hundreds AU.

Several models have been proposed to explain the X-ray emission from
protostellar jets, but the actual emission mechanism is still unclear.
\citet{bfr03} speculated on different mechanisms for the X-ray
emission from HH\,154: X-ray emission from the central star
reflected by a dense medium, X-ray emission produced when the stellar
wind shocks against the wind from the companion star, or produced in
shocks in the jet.  \citet{rnv02} derived a simple analytic model,
predicting X-ray emission originating from protostellar jets with the
observed characteristics.

Prompted by this recent detection of X-ray emission from HH objects,
we developed a detailed hydrodynamic model of the interaction between
a supersonic protostellar jet and the ambient medium, to explain the
mechanism causing the X-ray emission observed. Our model takes into
account optically thin radiative losses and thermal conduction effects. We
use the FLASH code (\citealt{for00}) with customized numerical modules
that treat optically thin radiative losses and thermal conduction
(\citealt{opr05}). The core of FLASH is based on a directionally split
Piecewise Parabolic Method (PPM) solver to handle compressible flows
with shocks (\citealt{cw84}). FLASH uses the PARAMESH library to handle
adaptive mesh refinement (\citealt{mom00}) and the Message-Passing
Interface library to achieve parallelization.

In a previous paper (\citealt{bop04}), we presented a first set of results
concerning a jet less dense than the ambient medium, with density contrast
$\nu = n_{\rm a}/n_{\rm j} = 10$ (where $n_{\rm a}$ is the ambient density
and $n_{\rm j}$ is the density of the jet) which emits X-rays in good
agreement with the X-ray emission observed in HH\,154 (\citealt{ffm02}).
\citet{bop04} have shown the validity of the physical principle on which
our model is based: a supersonic jet traveling through the ambient medium
produce a shock at the jet/ambient interaction front leading to X-ray
emission in good agreement with observations.  In the present paper, we
study the effects on the jet dynamics of varying the parameters, such
as the ambient-to-jet density ratio, $\nu = n_{\rm a}/n_{\rm j}$, and the Mach number, $M = v_{\rm j}/c_{\rm a}$,
through a wide range, to determine the range of parameters which can
give rise to X-ray emission consistent with observations.
Note that we use this definition for the Mach number to be able to compare the jet velocity with the ambient sound speed, to have information on how much the jet is supersonic.

The paper is structured as follow: Sect.~\ref{The model} describes
the model and the numerical setup; in Sect.~\ref{Results} we discuss the results of
our numerical simulations; finally Sect.~\ref{Discussion and conclusions}
is devoted to summary and conclusions. 
In Appendix ~\ref{synthesizing the X-ray
spectra} we discuss our method to synthesize X-ray emission from our
numerical simulations.

\section{The model}
\label{The model}

We model the propagation of a constantly driven protostellar jet through
an isothermal and homogeneous medium. We assume that the fluid is fully
ionized and that it can be regarded as a perfect gas with a ratio of
specific heats $\gamma = 5/3$. Also we assume a negligible magnetic field.

The jet evolution is described by the fluid equations of mass, momentum
and energy conservation, taking into account the effects of radiative
losses and thermal conduction

\begin{equation}
\frac{\partial \rho}{\partial t} + \nabla \cdot \rho \mbox{\bf v} = 0
\label{eq:massa-1}
\end{equation}

\begin{equation}
\frac{\partial \rho \mbox{\bf v}}{\partial t} +\nabla \cdot \rho
\mbox{\bf vv} + \nabla p = 0
\label{eq:momento-1}
\end{equation}

\begin{equation}
\frac{\partial \rho E}{\partial t} +\nabla\cdot (\rho E+p)\mbox{\bf v}
= - \nabla\cdot q - n_e n_H P(T)
\label{eq:en+r+c-1}
\end{equation}

\noindent
where $t$ is the time, $\rho$ the mass density, $\bf v$ the plasma
velocity, $p$ the pressure, $q$ the heat flux, $n_e$ and $n_H$ are the
electron and hydrogen density respectively, $P(T)$ is the optically thin
radiative losses function per unit emission measure (for the P(T) we use a functional form, which takes into account: free-free, bound-free, bound-bound and 2 photons emission, see
\citealt{rs77};
\citealt{mgv85}; \citealt{km00}), $T$ the plasma temperature, and

\begin{equation}
E = \epsilon +\frac{1}{2} |\mbox{\bf v}|^2
\end{equation}

\noindent
where $E$ is  the total energy and $\epsilon$ the specific internal
energy. We use the equation of state for an ideal gas
 
\begin{equation}
p=(\gamma-1)\rho\epsilon~.
\end{equation}

\noindent
Following \citet{db93}, we use an interpolation expression for the
thermal conductive flux of the form

\begin{equation}
q = \left(\frac{1}{q_{\rm spi}}+\frac{1}{q_{\rm sat}}\right)^{-1}
\label{eq:flusso} 
\end{equation}

\noindent
which allows for a smooth transition between the classical and saturated
conduction regime. In the above expression, $q_{\rm spi}$ represents the
classical conductive flux (\citealt{spi62})

\begin{equation}
q_{\rm spi} = - \kappa (T) \nabla T
\label{eq:flussospi}
\end{equation}

\noindent
where $\kappa (T) = 9.2\times10^{-7} T^{5/2}$ erg s$^{-1}$ K$^{-1}$ cm$^{-1}$ is the thermal conductivity. The
saturated flux, $q_{\rm sat}$, is (\citealt{cm77})

\begin{equation}
q_{\rm sat} = - sign(\nabla T) 5 \phi \rho c_{\rm s}^{3}
\label{eq:flussosat}
\end{equation}

\noindent
where $\phi\sim0.3$ (\citealt{1984ApJ...277..605G};
\citealt{1989ApJ...336..979B}, and references therein) and $c_{\rm s}$
is the isothermal sound speed.

\subsection{Numerical setup}
\label{sec:numerical setup}

We adopt a $2$-D cylindrical ($r, z$) coordinate system with the jet
axis coincident with the $z$-axis. For the different cases analyzed, we
have chosen different ranges for the radial and longitudinal dimensions of
the computational grid to follow in all cases the jet/ambient interaction
for at least 20-50 years: the computational grid size varies from $\approx
300$ AU to $\approx 600$ AU in the $r$ direction and from $\approx6000$
AU to $\approx 3\times10^{4}$ AU in the $z$ direction.

In the case of the jet less dense than the ambient medium (hereafter
``light jet'') which best reproduces observations, the integration
domain extends over $300$ AU in the radial direction and over $6000$
AU in the $z$ direction. In the case of the jet with the same initial
density as the ambient medium (hereafter ``equal-density jet'')
which best reproduces observations, the domain is $(r\times z) \approx
(600\times6000)$ AU. In this case, the radial axis is twice as large
than in the light jet case because the cocoon surrounding the equal
density jet has a radial extension greater than in the light jet.
The dimension of the computational domain in the case of the jet
denser than the ambient (hereafter ``heavy jet'') which best reproduces
observations is $(r\times z) \approx (700 \times 27000)$ AU.

\begin{figure}[!t]
\centerline{\psfig{figure=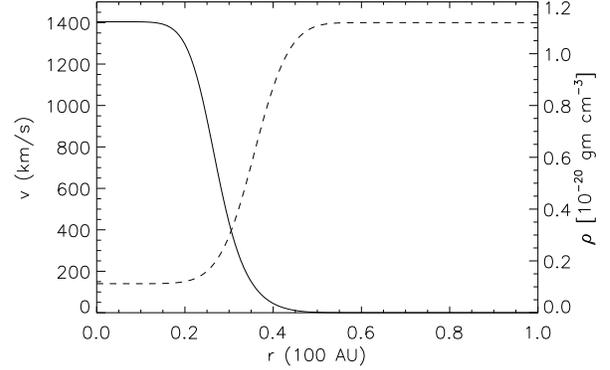,width=8cm}}
\caption{Initial jet velocity (continuous line) and density
(dashed line) as a function of the distance from the axis, $r$, for the light jet case discussed in Sect. \ref{Hydrodynamic evolution} with parameters: $M = 300$, $\nu=10$, $n_{j} = 500$ cm$^{-3}$, $T_{a} = 10^{3}$ K, $r_{j} = 30$ AU, $w = 4$ (see Tab. \ref{tab_mod}).}
\label{fig1}
\end{figure}

In all the cases, the initial jet velocity is along the
$z$ axis, coincident with the jet axis, and has a radial profile of
the form

\begin{equation}
V(r) = \frac{V_0}{\nu \cosh(r/r_j)^{w}-(\nu-1)}
\end{equation}

\noindent
where $V_{0}$ is the on-axis velocity, $\nu$ is the ambient to
jet density ratio, $r_{\rm j}$ is the initial jet radius and $w \bf= 4$
is the steepness parameter for the shear layer (as an example, see continuous line
in Fig.~\ref{fig1}, for the light jet case discussed in Sect. \ref{Hydrodynamic evolution} with parameters in Tab. \ref{tab_mod}), to have a smooth
transition of the kinetic energy at the interface between the jet and
the ambient medium.

The density variation in the radial direction (dashed line in
Fig.~\ref{fig1}) is

\begin{equation}
\rho(r) = \rho_j\left(\nu-\frac{\nu-1}{\cosh(r/r_j)^{w}}\right) 
\end{equation}

\noindent
where $\rho_{\rm j}$ is the jet density (\citealt{bmf94}).

Reflection boundary conditions are imposed along the jet axis, inflow
boundary conditions at $z = 0$ and $r \leq {r_{\rm j}}$ and outflow
boundary conditions elsewhere.

The maximum spatial resolution achieved in the best light jet case, (in both $r$ and $z$ directions), 
is $\approx 1.3$ AU according to the PARAMESH methodology, using $4$
refinement levels, corresponding to covering the jet radius with $25$
points at the maximum resolution.  The spatial resolution achieved in
the equal-density case is half the one in the light jet case.  In the
best heavy jet model, the spatial resolution achieved is $8$ times lower
than in the light jet case.

Our choice of different spatial resolution for the three cases, aimed
at reducing computational cost, was necessary because the thermal
conduction is solved explicitly in FLASH and, therefore, a time-step
limiter depending on density, $\rho$, temperature, $T$, and spatial
resolution, $\Delta x$, is required to avoid numerical instability (see,
for instance, \citealt{opr05}). Stability is guaranteed for $\Delta t <
0.5~ \Delta x^2/D$, where $D$ is the diffusion coefficient, related to
the conductivity, $\kappa$, and to the specific heat at constant volume,
$c_v$, by $D = \kappa (T)/(\rho c_v)$. In models characterized by high
values of temperature (as, for instance, in the heavy jet case),
therefore, a lower spatial resolution was required to avoid a very small
time-step, $\Delta t$.

\subsection{Time scales}
\label{Time scales}

Condensations of plasma, due to radiative cooling effects, can
become thermally unstable; however the presence of thermal conduction can
prevent such instabilities. By comparing the radiative, $\tau_{\rm rad}$,
and thermal conduction, $\tau_{\rm cond}$, characteristic times

\begin{equation}
\tau_{\rm rad} = \frac{p}{(\gamma-1) n_e n_H P(T)}\approx 2.5\times10^{3}
\frac{T^{3/2}}{n}
\end{equation}

\begin{equation}
\tau_{\rm cond}\approx \frac{p}{\gamma-1} \frac{7}{2} \frac{l^{2}}{\kappa (T) T}\approx1.5\times10^{-9} \frac{nl^{2}}{T^{5/2}}~,
\end{equation}

\noindent
where $l$ represents the characteristic length of temperature variations,
we can infer which of the two competing processes dominates during the jet/ambient interaction. From the condition

\begin{equation}
\left(\frac{\tau_{\rm rad}}{\tau_{\rm cond}}\right)^{1/2} = 1
\label{eq:trad/tcond-2}
\end{equation}

\noindent
we can derive the cutoff length scale for instability, $l_{\rm F}$
(\citealt{fie65}), which indicates the maximum length

\begin{equation}
l_{\rm F}\approx 1.3\times10^{6} \frac{T^{2}}{n}
\label{field_len}
\end{equation}

\noindent
over which thermal conduction dominates over radiative effects,in the
classical conduction regime. An analogous estimate in the saturation
regime leads to

\begin{equation}
\left(l_{\rm F}\right)_{\rm sat}\approx 3\times10^{7}
\frac{T^{2}}{n}~,
\end{equation}

\noindent
which is one order of magnitude longer than the characteristic length
in the classical regime. As discussed later in Sect. \ref{Hydrodynamic
evolution}, the comparison between the classical Field length (the
shortest characteristic length) and the size of the region behind the
shock at the head of the jet will allow us to determine if this region
is thermally stable or not.

In order to verify our hypothesis of a fully ionized gas, we computed the ionization time scale of the most relevant elements in the X-ray spectrum of a
shocked plasma at $T = 3.4\times10^{6}$ K, assuming a post-shock density of about $10^{4}$ cm$^{-3}$ (the light
jet case). As an example, we derived that the ionization time scale for C and O is 1 to 2 orders of magnitudes smaller than the radiative and thermal conduction time scales so that the plasma can be considered in equilibrium.

\subsection{Parameters}
\label{Parameters}

Our model solutions depend upon several physical parameters as, for
instance, the jet and ambient temperature and density,  the jet
velocity and its radius.
With the aim to reduce the number of free parameters in our exploration of the parameter space, we fixed the jet radius to $r_{j} \approx 30$ AU, according to \citet{ffm02} who found this characteristic linear scale, from the X-ray thermal fit, and according to \citet{fld05} who showed HST images of the internal knots of HH 154 with dimension $r\approx30$ AU at the base of the jet\footnote{In \citet{fld05}, on page 993 the authors discuss the working surface. The radius quoted is the one of the elongated Mach disk (probably representative of the jet) and it is $0.3/2''\approx30$ AU. The separation between Mach disk and working surface is $0.6"$ or $4$ times this  $\approx100$ AU (M. Fridlund, private communication).}. However detailed simulations with different $r_{j}$ values are not necessary since we can predict the effects of varying the jet radius from the model results we obtained so far. In fact we expect the X-ray emitting region to grow in size as $r_{j}$ grows. 
Since the X-ray luminosity is defined as $L_{\rm X} = n^{2} V P(T)$, it depends on the cube of the radius. This means that, $L_{\rm X}$ being constrained from observations, a jet with a greater radius needs a lower density in order to reproduce observations.
We impose an initial jet length $z_{j} = 300$ AU to avoid the ejected plasma that travels back inside the boundary during the jet evolution. This choice of a non-zero initial jet length allows us to obtain an unperturbed boundary surface at $z = 0$.
In all our simulations, we model a jet with initial density
and temperature $n_{\rm j} = 500$ cm $^{-3}$ and $T_{\rm j} = 10^{4}$
K respectively, according to the values derived from observations
(\citealt{fl98} and \citealt{ffm02}). 
The density and temperature of the ambient medium, $n_{\rm a}$ and $T_{\rm
   a}$ respectively, are derived from the choice of the ambient-to-jet density contrast, $\nu$ and
   from the hypothesis of initial pressure balance between the ambient
   medium and the jet.
We are left, therefore, with
two non-dimensional control parameters: the jet Mach number, $M$, and
the ambient-to-jet density contrast, $\nu$. 
For a more extended exploration of the parameter space, see Sect. \ref{Varying the jet density parameter}, concerning the variation of the initial jet density, $n_{\rm j}$.
In our simulations we account for a wide jet/ambient parameters range shown in Tab. \ref{tab:range}.

\begin{table*}[!t]
\caption{Range of parameters used in our numerical model (column 2) compared with typical outflow parameters (column 3 and 4) shown in \citet{br02}, Tab. 1.
$\nu$ is the ambient to jet density contrast; $M$ is the Mach number; $T_{a}$ and $n_{a}$ are the ambient temperature and density, respectively; $v_{\rm j}$ is the initial jet velocity; $v_{\rm sh}$ is the shock velocity of the knots inside the jet; $\dot{M}$ is the mass loss rate; $L_{\rm mech}$ is the mechanical luminosity.}\label{tab:range}
\par
\begin{tabular}{lcccc}
\hline
\hline
Parameter      & Model                    & Low Mass$^{a}$       & High Mass$^{a}$ & Units \\
\hline
$\nu$          & $0.01\div300$            & $-$                  & $-$                  & $-$\\
$M$            & $1\div1000$              & $-$                  & $-$                  & $-$\\
$T_{a}$        & $30\div10^{6}$           & $-$                  & $-$                  & K\\
$n_{a}$        & $5\div10^{5}$            & $-$                  & $-$                  & cm$^{-3}$\\
$v_{\rm j}$        & $85\div8500$         & $-$                  & $-$                  & km s$^{-1}$ \\
$v_{\rm sh}$        & $100\div2000$             & $100\div300$         & $100\div1000$        & km s$^{-1}$ \\
$\dot{M}$      & $10^{-10}\div10^{-8}$    & $10^{-9}\div10^{-5}$ & $10^{-6}\div10^{-2}$ & $M_{\odot}$ yr$^{-1}$ \\
$L_{\rm mech}$ & $6.7\times10^{-5}\div67$ & $0.001\div1$         & $0.1\div1000$        & $L_{\odot}$ \\
\hline
\hline
\noalign{\smallskip}
\multicolumn{4}{l}{$^a$ \citet{br02}}
\end{tabular}
\end{table*}

In Sect. \ref{Results},
we discuss the results derived from the exploration of the parameters
space defined by $M$ and $\nu$.

\section{Results}
\label{Results}

\subsection{Exploration of the parameter space}

We performed a wide exploration of the parameter space defined by two
free parameters: the jet Mach number, $M = v_{\rm j}/c_{\rm a}$, and the
ambient-to-jet density ratio, $\nu = n_{\rm a}/n_{\rm j}$ (see Sect.
\ref{Parameters}). The aim is to determine the range of parameters
leading to X-ray emission from protostellar jets in agreement with the observations.

We first analyzed adiabatic hydrodynamic models, i.e. without thermal
conduction and radiative losses. Then, for the most promising cases
(i.e. those which reproduce the values of jet velocity, temperature
and luminosity of the X-ray source derived from the observations),
we performed more realistic simulations in which we have taken into
account thermal conduction and radiative losses effects. By comparing
these models with those without thermal conduction and radiative cooling,
we explored how the presence of these physical processes affects the
jet/ambient system evolution. We found that,
in general, models with thermal conduction and radiation reach lower
temperatures (up to $5$ times lower than those achieved in the adiabatic
cases). We also found that thermal conduction smooths the structures
(well visible in the pure hydrodynamic cases) in the density and
temperature distributions.

\begin{figure}[!t]
\centerline{\psfig{figure=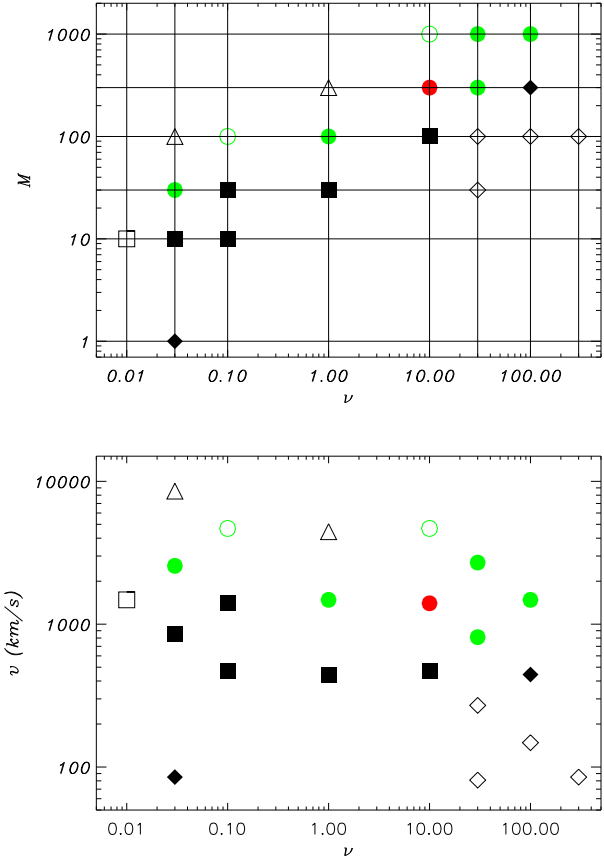,width=8cm}}
\caption{Exploration of the parameter space: jet Mach number, $M$
(upper panel) and initial jet velocity (lower panel) as a function of
the ambient-to-jet density contrast, $\nu$.  Green and red dots refer to
models consistent with observations for X-ray luminosity, shock velocity
and fitting temperature values (see text); diamonds refer to models which
cannot reproduce observations; squares refer to models in good agreement
with observations for shock velocity but not for X-ray luminosity values;
triangles refer to models with too high ($T\geq 10^{7}$ K, one order
of magnitude higher than observed) temperature. Empty symbols refer to
pure hydrodynamic simulations; filled symbols refer to models in which
thermal conduction and radiative losses effects are taken into account.}
\label{fig:griglia}
\end{figure}

In the following subsections, we discuss the models (shown in
Fig.~\ref{fig:griglia}) in which both radiative losses and thermal
conduction are taken into account. In Fig.~\ref{fig:griglia} green and
red dots refer to those cases with X-ray luminosity $L_{X} > 10^{28}$
erg s$^{-1}$, shock front velocity $v_{sh} > 100$ km s$^{-1}$ and fitting
temperature $T < 10^{7}$ K, consistent with observations. We have chosen
$L_{X}$ one order of magnitude lower than the minimum value observed (see
Tab.~\ref{tab:obs}) to take into account fainter sources not detected so
far; the red dot refers to the representative case of HH\,154 discussed
in \citet{bop04}. Squares show cases with velocity
in the range of values observed, but with $L_{X} < 10^{28}$ erg s$^{-1}$.
Diamonds mark the cases with velocity and X-ray luminosity not consistent
with observations. Triangles mark cases with temperatures higher than
$10^{7}$ K. The lower panel of Fig. \ref{fig:griglia} show the initial
velocity assumed in our simulations vs. the density contrast.

From our exploration of the parameter space, we derived that the models in
agreement with observations are included in a well constrained region. In
the following sections, we discuss in details the ``best-fit'' models,
i. e. those models which reproduce X-ray luminosity and shock front
speed values as close as possible to those observed, in the cases of
light, equal-density and heavy jets (see Tab.  \ref{tab_mod}).

   \begin{table}[!t]
   \caption{Summary of the initial physical parameters characterizing the
   ``best-fit'' models in the cases of light, equal-density and heavy
   jets: ambient-to-jet density contrast, $\nu$, jet Mach number, $M$,
   initial jet velocity, $v_{\rm j}$, ambient density and the temperature,
   $n_{\rm a}$, and $T_{\rm a}$, respectively.  In all the models, the initial jet density and
   temperature are $n_{\rm j} = 500$ cm $^{-3}$ and $T_{\rm j} = 10^{4}$
   K, respectively.}
   \label{tab_mod}
   \begin{tabular}{lccccc}
   \hline
   \hline
   model & $\nu$  & $M$ & $v_{\rm j}$ & $n_{\rm a}$ & $T_{\rm a}$ \\
         & & & [km s$^{-1}$] & [cm$^{-3}$] & [$10^4$ K]\\
   \hline
   light         & 10   & 300 & 1400 & 5000 & 0.1 \\
   equal-density & 1    & 100 & 1500 & 500  & 1   \\
   heavy         & 0.03 & 30  & 2500 & 17   & 30  \\
   \hline
   \end{tabular}
   \end{table}

\subsection{Hydrodynamic evolution}
\label{Hydrodynamic evolution}

\begin{figure}[!t]
\centerline{\psfig{figure=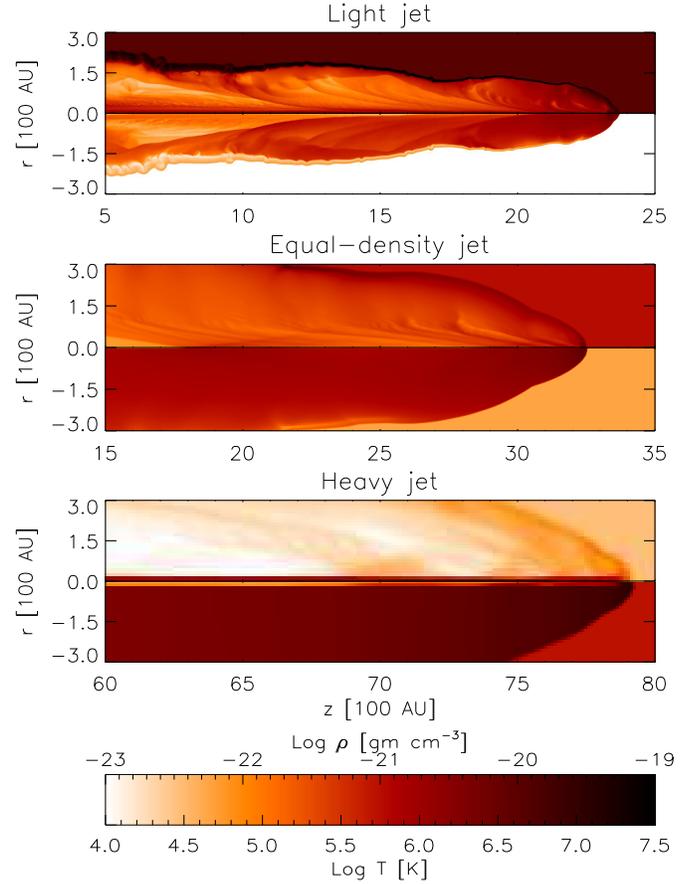,height=12cm}}
\caption{Mass density (upper semi-panels) and temperature (lower
semi-panels) bi-dimensional cuts in the $r-z$ plane after $20$ years
since the beginning of the jet/ambient interaction for the best cases
of light (upper panels), equal-density (middle panels) and heavy jet
(lower panels).}
\label{fig:mappe-20yr}
\end{figure}

In Fig.~\ref{fig:mappe-20yr}, we show the mass density and temperature
distributions $20$ years since the beginning of the jet/ambient medium
interaction for the three best-fit models in Tab. \ref{tab_mod}. The light
jet case is the one which better reproduce the physical parameters derived
from observations by \citet{fl98} and \citet{ffm02} for the HH\,154
protostellar jet; its properties have been discussed in \citet{bop04}.

\begin{figure}[!t]
\centerline{\psfig{figure=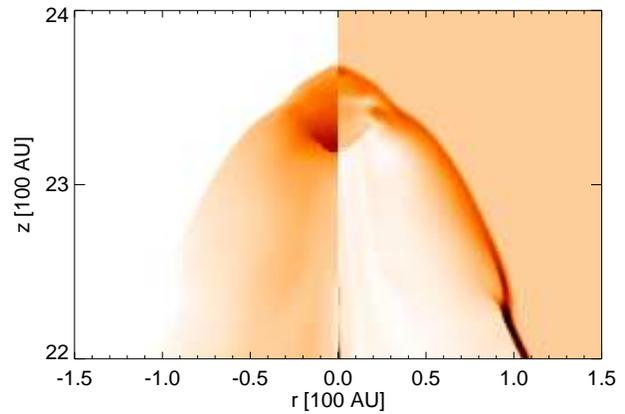,width=7cm}}
\centerline{}
\centerline{}
\caption{An enlargement of the post shock region in the light jet case
of temperature (left) and density (right) in linear scale, about $20$
years since the beginning of the jet/ambient interaction. Note
the hot and dense region just behind the shock front.}
\label{fig:zoom-light}
\end{figure}

In all the cases, at the head of the jet there is clear evidence of a
shock front due to the plasma propagating supersonically along the jet
axis. Just behind the shock front there is a localized hot and dense
blob (see, for instance, the enlargement in Fig.~\ref{fig:zoom-light}
for the light jet case).

\begin{figure}[!t]
\centerline{\psfig{figure=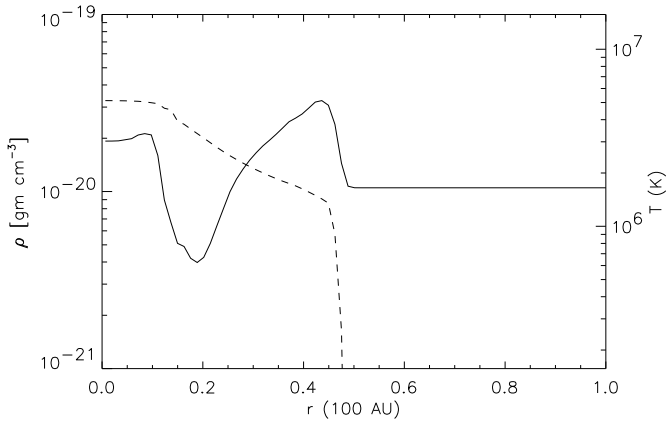,width=8cm}}
\caption{Density (continuous line) and temperature (dashed line)
profile along the radial direction at $z\approx 5000$ AU corresponding
to the blob position at $t = 40$ yr ($M = 300$, $\nu = 10$).}
\label{fig:dens-radial-222}
\end{figure}

The light jet is enveloped by a cocoon with temperature $T \approx
7\times10^{5}$ K, almost uniform due to the thermal conduction diffusive
effect; nevertheless the cocoon temperature is not constant in time
but decreases as the evolution goes on, leading to the formation of a
cool and dense external envelope.  Fig.~\ref{fig:dens-radial-222}
shows cuts of the density (continuous line) and temperature (dashed
line) along the radius at $z\approx5000$ AU, corresponding to the blob
position 40 years since the beginning of the jet/ambient interaction:
the hot (few millions degrees) and dense blob is evident for $r<10$
AU. The density decreases moving away from the jet axis along the radial
direction and, then, it increases again at the position corresponding
to the external part of the cocoon. On the other hand, the temperature
monotonously decreases, moving away from the jet axis along the radial
direction. The blob, therefore, is expected to be an X-ray source; in
Sect.~\ref{Spectral analysis}, we will show that
the X-ray source has luminosity and spectral characteristics consistent
with those observed.

The central panel in Fig.~\ref{fig:mappe-20yr} shows 2-D sections in the
$(r,z)$ plane of the mass density and temperature distributions for the
best-fit equal-density model (see Tab. \ref{tab_mod}). The interaction
between the protostellar jet and the ambient medium causes the presence
of a dense and hot cocoon ($n\approx1400$ cm$^{-3}$; $T\approx2\times
10^ {6}$ K) surrounding the jet. Once again the cocoon is almost uniform
for the presence of the thermal conduction but its temperature decreases
with time. The cocoon becomes gradually cool
and dense with time as in the light-jet case. Also in
this case, the post-shock region is a hot and dense blob from which the
X-ray emission originates (see Sect. \ref{Spatial distribution of X-ray
emission} for more details).

In the heavy-jet case (lower panel in Fig.  \ref{fig:mappe-20yr}),
the jet is surrounded by a cocoon well smoothed by the effects of the
thermal conduction and with a radial extension larger than in the other
two cases. The cocoon has temperature of a few millions degrees and
density lower than that of the jet.

For the three best-fit models discussed above, we analyzed the thermal
stability of the hot and dense blob localized behind the shock by
comparing the size of the blob with the Field length Eq. \ref{field_len}.
In the light jet case, the values obtained for the average blob
temperature, $T\approx3.4\times10^{6}$ K, and density, $n\approx6500$
cm$^{-3}$, lead to $l_{\rm F} \sim 100$ AU. Since the blob size,
almost equal to twice the initial jet radius $r_{\rm j}\approx 30$ AU
(see Fig. \ref{fig:zoom-light}), is smaller than $l_{\rm F}$, it turns
out that the blob is thermally stable. In the equal-density jet case,
the density and temperature of the blob at the head of the jet are
$n\approx1700$ cm$^{-3}$ and $T\approx4.3\times10^{6}$, respectively,
leading to $l_{\rm F}\approx 10^3$ AU. Also in this case, therefore,
the blob is thermally stable, being its size $\sim100$ AU, i.e. $10$ times
smaller than the Field length. In the heavy-jet case, the temperature
and density of the blob are $T\sim10^{7}$ K and $n\sim10^{2}$ cm$^{-3}$,
leading to $l_{\rm F}\approx 10^{4}$ AU. Since the blob behind the shock
front extends over about $100$ AU, also in this case it is thermally
stable.

\begin{figure}[!t]
\centerline{\psfig{figure=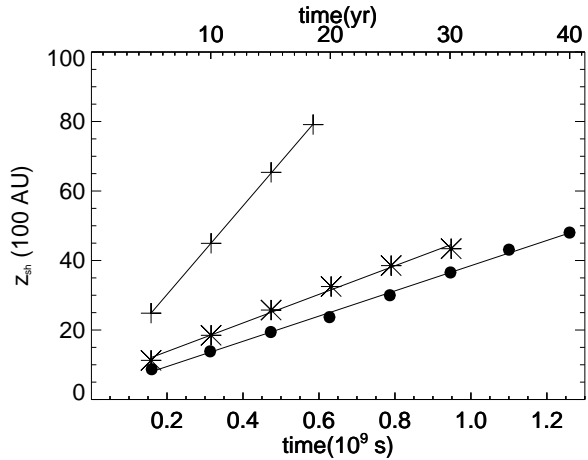,width=8cm}}
\caption{Shock front position vs. time. Dots mark the light jet case,
stars the equal-density jet and crosses the heavy jet case. The lines
refer to the best-fit from which we derive the average shock speed:
$v_{\rm sh}\approx 500$ km s$^{-1}$, $\approx 600$ km s$^{-1}$, $\approx
1900$ km s$^{-1}$ for the light, equal-density and heavy jet case,
respectively.} \label{fig:zVSt-per3} \end{figure}

The position of the shock front as a function of time for
the three best-fit models in Tab. \ref{tab_mod} is shown in
Fig. \ref{fig:zVSt-per3}. For the light jet case, we derived an average
shock velocity $v_{\rm sh}\approx 500$ km s$^{-1}$, about $3$ times lower
than the initial jet velocity. This shock velocity is in good agreement
with observed speeds in HH objects and in particular with that derived
from HH\,154 data. Taking into account the jet inclination $\approx
45$ degrees (\citealt{fl98}), $v_{\rm sh}\approx 500$ km s$^{-1}$
corresponds to a proper motion of $\approx 350$ km s$^{-1}$, which,
at the distance of HH\,154, can be measured with well time-spaced
\emph{Chandra} observations.

In the equal-density jet scenario, we deduced an average shock velocity
$v_{\rm sh}\approx 600$ km s$^{-1}$, slightly greater than the value
observed in HH\,154 (\citealt{fl98}; \citealt{ffm02}), but consistent
with values observed in other HH objects (see Tab.~\ref{tab:obs}). For
the heavy jet case, the average value of the shock speed is $v_{\rm
sh}\approx 1900$ km s$^{-1}$ which is too high with respect to the HH
shock front velocities observed (cf. Tab.~\ref{tab:obs}).

\subsection{Emission measure distribution vs. temperature}
\label{Emission measure distribution vs. temperature}

\begin{figure}[!t]
\centerline{\psfig{figure=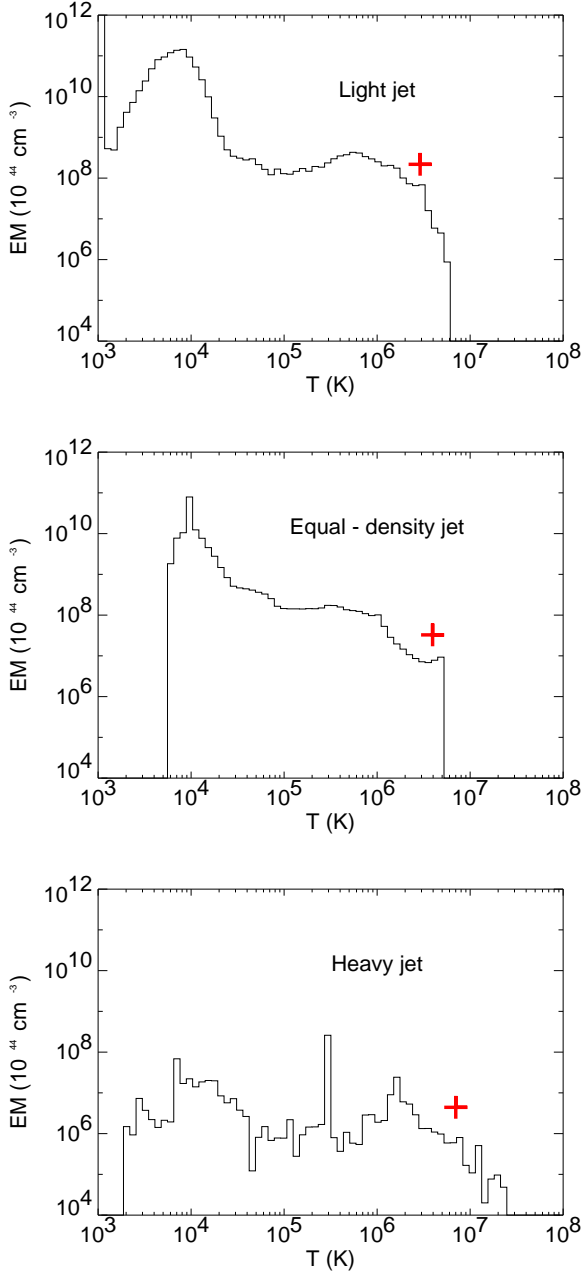,width=8cm}}
\caption{Emission measure, $EM$, as a function of the
temperature, $T$ for the three models discussed, $20$ years since the
beginning of the jet/ambient interaction. The cross superimposed on each
panel marks the best-fit temperature and emission measure values derived
from our simulated absorbed spectra.}
\label{fig:EMVST-per3}
\end{figure}

We derived the distribution of emission measure vs. temperature, $EM(T)$,
in the temperature range [$10^{3}-10^{8}$] K at different stages of the
evolution of the jet/ambient system (see Appendix \ref{synthesizing the
X-ray spectra} for more details). Fig. \ref{fig:EMVST-per3} shows the $EM(T)$ for the three
best-fit models in Tab. \ref{tab_mod}, after $20$ years since the beginning of
the jet/ambient interaction.

In all the cases, we found that the shape of the $EM(T)$ is characterized
by two bumps, and does not change significantly during the system
evolution. The relative weight of the bumps is different in the three
cases. In the light jet case (upper panel in Fig.~\ref{fig:EMVST-per3}),
the bumps are quite broad, the first centered at temperature $T\sim
10^{4}$ K with $EM\sim 10^{55}$ cm$^{-3}$, and the second one centered
at $T\sim10^{6}$ K with $EM\sim 10^{52}$ cm$^{-3}$, about three orders
of magnitude lower than the previous one; the $EM$ decreases rapidly
above few millions degrees.

In the equal-density jet case (middle panel in Fig.~\ref{fig:EMVST-per3}),
the first bump is centered at $T\sim10^{4}$ K with $EM\geq10^{54}$
cm$^{-3}$, whereas the second bump is centered at $T\sim10^{6}$ K with
$EM\sim10^{52}$ cm$^{-3}$, as in the light jet case. In the heavy jet
case (lower panel in Fig.~\ref{fig:EMVST-per3}), the $EM(T)$ distribution
appears flat with two weak peaks: the first centered at $T\sim10^{4}$
K and the second one at a few millions degrees. Note that, in the heavy
jet case, the EM at temperature up to a few million degrees is two
orders of magnitude lower than in the light jet case.

On the base of the $EM(T)$ distribution, we expect a bright ($L_{\rm
X} > 10^{28}$ erg s$^{-1}$) X-ray source, whose soft component (due to the
cocoon) could be suppressed by the strong interstellar medium absorption
(\citealt{ffm02}). We expect also that the X-ray emission decreases as
the ambient-to-jet density ratio, $\nu$, decreases, leading to a brighter
X-ray emission in the light jet case.

\subsection{X-ray emission}

From the $EM(T)$ distributions and the MEKAL spectral code, we synthesized
the focal plane spectra as predicted to be detected with the instruments
on board XMM-\emph{Newton} and \emph{Chandra} (see
Appendix \ref{synthesizing the X-ray spectra} for details), taking into account the
interstellar absorption. To compare our numerical models with experimental
data concerning HH\,154 (the closest and best studied jet emitting in
the X-ray band), we assumed a distance of 150 pc (as HH\,154 that is
located in the L1551 cloud in the Taurus star-forming region) and an
interstellar absorption column density $N_{\rm H} = 1.4\times10^{22}$
cm$^{-2}$ (\citealt{ffm02}). Our model results can be generalized to
account for the other HH objects observations by considering different
values for the distance and the interstellar absorption.

\subsubsection{Spatial distribution of the X-ray emission}
\label{Spatial distribution of X-ray emission}

Assuming the jet propagating perpendicularly to the line of sight, from
the numerical simulations we derived X-ray images of the jet/ambient
system as predicted to be observed with \emph{Chandra}/ACIS-I (see
Appendix \ref{synthesizing the X-ray spectra}) that allows us to pinpoint
the X-ray emission, thanks to its high spatial resolution.
For all the models in Tab. \ref{tab_mod}, we derived the evidence
that most of the X-ray emission produced during the jet/ambient medium
interaction originates from a very compact region localized at the head
of the jet, just behind the shock front.

\begin{figure}[!t]
\centerline{\psfig{figure=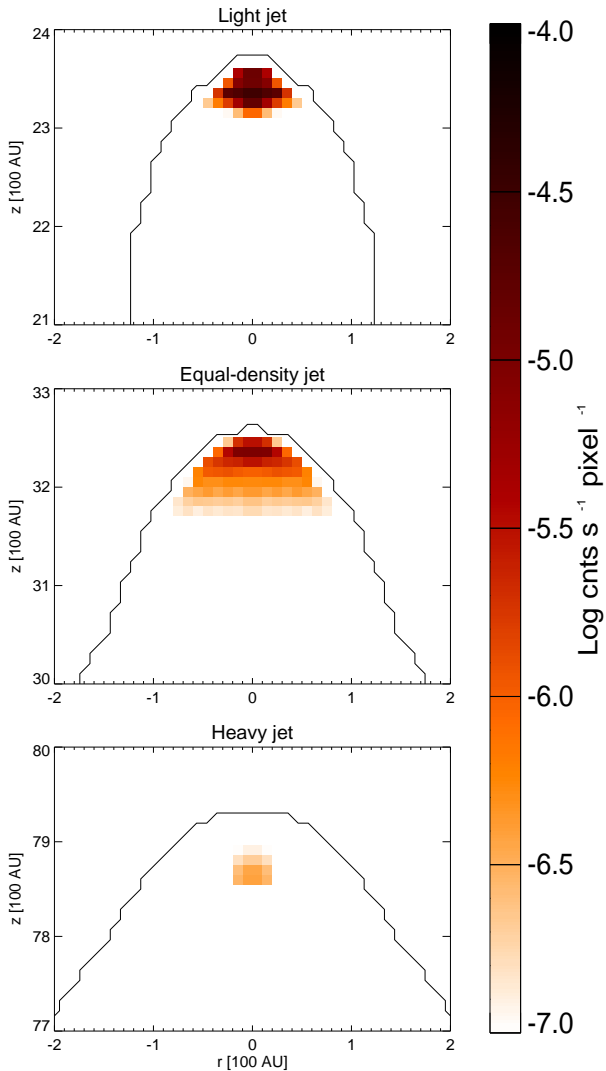,width=8cm}}
\caption{Synthesized X-ray emission, in logarithmic scale, as observed
with ACIS-I, for the three cases examined, $20$ years since the beginning
of the jet/ambient interaction. At a distance $D \approx 150$ pc, 100
AU correspond to about 0.7 arcsec}
\label{fig:mappeX-20yr}
\end{figure}

Fig.~\ref{fig:mappeX-20yr} shows an enlargement of the X-ray images as
predicted to be detected with \emph{Chandra}/ACIS-I of the head of the jet
where most of the X-ray emission originates, 20 years since the beginning
of the jet/ambient interaction. Note that the spatial resolution of the
X-ray images in Fig.~\ref{fig:mappeX-20yr} is 6 times better than that
of \emph{Chandra}/ACIS-I. In all the three cases analyzed, a comparison
between the X-ray emitting region and the temperature and density maps in
Fig.~\ref{fig:mappe-20yr} shows that the X-ray source is coincident with
the hot and dense blob discussed in Sect. \ref{Hydrodynamic evolution}.

We found that even with \emph{Chandra}'s spatial resolution, the X-ray
emitting region cannot be spatially resolved (the spatial resolution
of the synthesized X-ray images in Fig. \ref{fig:mappeX-20yr} is 6
times better than that of \emph{Chandra}), and it will be detected as
a point-like source. Also, significant X-ray emission is visible only
from the hot and dense blob behind the shock front, as softer cocoon
emission is extinguished by the the strong interstellar absorption.

As discussed in Sect. \ref{Hydrodynamic evolution}, the X-ray emitting
region for the three cases examined is thermally stable, and therefore
the X-ray emission is detectable continuously during the 20-50 years
analyzed. We found also that there are no significant variations of the
X-ray source morphology during the evolution: the source size varies
by $\pm 25 \%$, always below the spatial resolution achievable with
\emph{Chandra}, and there is no disappearance of the X-ray emitting
region, according to the stability analysis. On the other hand, for some
of the cases shown in Fig.~\ref{fig:griglia}, our analysis predicts a
transient behaviour of the X-ray source, which extinguishes, few years
since the beginning of the interaction between the protostellar jet and
the ambient medium, because of the radiative cooling which dominates
over the thermal conduction effects.

From Fig. \ref{fig:zVSt-per3}, the X-ray source, coincident with the hot
and dense blob discussed in Sect. \ref{Hydrodynamic evolution}, has a
proper motion of $\sim0.7$ arcsec/yr, $\sim0.8$ arcsec/yr, and $\sim2.7$
arcsec/yr in the light, equal-density and heavy jet case respectively,
assuming the jet axis to be perpendicular to the line of sight. In
addition, we found that the intensity of the X-ray source decreases
about one order of magnitude as the ambient-to-jet density contrast,
$\nu$, decreases.  Note that the heavy jet case has the higher shock
front speed and the lower X-ray emission.

\subsubsection{Spectral analysis}
\label{Spectral analysis}

We derived the synthesized focal plane spectra as they would be detected
with XMM-\emph{Newton}/EPIC-pn, characterized by an high effective area,
with the aim to compare our model results with published data and, in
particular, with those concerning HH\,154 (\citealt{ffm02}).

We considered two different levels of count statistics in the $[0.3-10]$
keV band: in the low statistics case we have chosen an exposure time
so as to obtain about $100$ total photons for each spectrum whereas,
in the high statistics case, we imposed about $10^{4}$ counts for each
spectrum. Albeit the latter case is unrealistic, given the low photon
counts so far collected from these sources, it can help us to pinpoint
some fundamental features of the predicted spectra. The spectral bins
are grouped together to have at least 10 photons in the low count
statistics case and 20 photons in the other case.

For the models of light and equal-density jet, the synthesized spectra
are well described as the emission from an optically thin plasma at a
single temperature, even in the high count statistics case. This result
is due to the strong interstellar absorption which suppresses the soft
emission originating from the cooler plasma component in the cocoon. The
best fit parameters derived from our simulations are shown in
Tab.~\ref{tab:fit}.

\begin{table*}[!t]
\caption{Best-fit parameters for the EPIC-pn simulated X-ray spectra
obtained in the low and high statistics cases, respectively, for the
three best-fit models of Tab. \ref{tab_mod}, about 20 yr since the beginning
of the jet/ambient interaction.}
\label{tab:fit}
\par
\begin{tabular}{lcccccccc}
\hline
\hline
Model & counts & $N_{\rm H} \pm \Delta N_{\rm H}$& $T_1 \pm \Delta T_1$& $EM_1 \pm \Delta EM_1$& 
                                         $T_2 \pm \Delta T_2$& $EM_2 \pm \Delta EM_2$& 
                                         $\chi^{2}$ & Prob.$^a$ \\
   & & $(10^{22}$ cm$^{-2})$           & ($10^{6}$ K)    & ($10^{52}$ cm$^{-3})$&
                                         ($10^{6}$ K)    & ($10^{52}$ cm$^{-3})$&  \\
\hline
{\em Light}
 & 102    & $1.4  \pm 0.3$  & $2.9 \pm 1.2$ & $2.5 \pm10.9$ & $-$ & $-$ & 0.44 & 0.88\\
 & 10317  & $1.39 \pm 0.02$ & $2.9 \pm 0.1$ & $2.2 \pm 0.6$ & $-$ & $-$ & 0.73 & 1.00\\ 
\hline
{\em Equal-density}
 & 71     & $1.3  \pm 0.4$  & $3.1 \pm 3.8$ & $0.4 \pm 3.5$ & $-$ & $-$ & 0.49 & 0.75\\
 & 10134  & $1.38 \pm 0.02$ & $3.9 \pm 0.2$ & $0.33\pm 0.09$& $-$ & $-$ & 0.72 & 1.00\\ 
\hline
{\em Heavy}
 & 92     & $1.4  \pm 0.2$  & $6.8 \pm 1.7$ & $0.9 \pm 3.4$ & $-$ & $-$ & 0.39 & 0.93\\
 & 9811   & $1.40 \pm 0.04$ & $3.0 \pm 0.1$ & $0.2 \pm 0.2$ & $11.8 \pm 0.4$ & $0.016 \pm 0.002$ & 0.73 & 0.99\\ 
\hline
\noalign{\smallskip}
\multicolumn{8}{l}{$^a$ Null hypothesis probability.}\\
\end{tabular}
\end{table*}

The heavy jet model which best fits HH observations show more structured
spectra than those obtained in the light jet and equal-density jet cases:
in the high statistics case, the spectra are well described by a two
temperature plasma emission, with the best fit parameters reported in
Tab.~\ref{tab:fit}.

The spectral analysis reflects the structure of the $EM(T)$ distribution:
the spectra are sensitive to the high temperature portion of the $EM(T)$
(see Fig.~\ref{fig:EMVST-per3}), being the softer component suppressed
by the interstellar medium absorption. On the other hand, in the
heavy jet model, the $EM(T)$ distribution is characterized by a bump at
high temperatures broader than those in the other two cases (see
Fig. \ref{fig:EMVST-per3}), implying that more than a single temperature
contributes to the emission.

\begin{figure}[!t]
\centerline{\psfig{figure=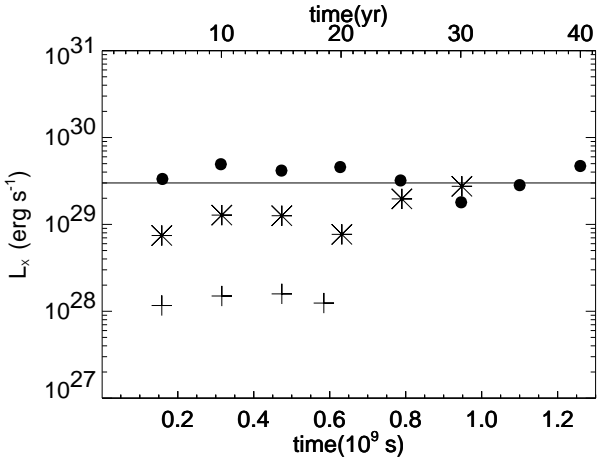,width=8cm}}
\caption{X-ray luminosity evolution as a function of time. Dots mark
the light jet case, stars the equal-density jet and crosses the heavy
jet case. The line superimposed on this figure marks $L_{\rm X} =
3\times10^{29}$ erg s$^{-1}$, the value observed in HH\,154 (\citealt{ffm02}).}
\label{fig:LX-isot-per3}
\end{figure}

Fig. \ref{fig:LX-isot-per3} shows the evolution of the X-ray luminosity,
$L_{\rm X}$, in the $[0.3-10]$ keV band, derived from the isothermal
components fitting the spectra in the high statistics case. The light
jet case (dots in Fig.~\ref{fig:LX-isot-per3}) shows $L_{\rm X}$
values in good agreement with those observed in HH\,154 ($L_{\rm
X} = 3\times10^{29}$ erg s$^{-1}$, see \citealt{ffm02}). In the
equal-density jet case, $L_{\rm X}$ ranges between $7.4\times10^{28}$
and $2.7\times10^{29}$ erg s$^{-1}$, in general below the luminosity
observed in HH\,154, although its values are consistent with those
detected in other HH objects (see Tab.~\ref{tab:obs}). On the other
hand, in the heavy jet case, the $L_{\rm X}$ values (crosses in
Fig.~\ref{fig:LX-isot-per3}) are at least one order of magnitude lower
than those observed so far in HH objects (Tab.~\ref{tab:obs}) and in
HH\,154 in particular.

\subsection{Varying the jet density parameter}
\label{Varying the jet density parameter}

As an extension of the exploration of the parameter space, we have varied the value of the initial jet density, $n_{\rm j}$, so far fixed to the value derived from \citet{fl98} for HH 154 (namely $n_{\rm j} = 500$ cm$^{-3}$). In particular, we performed the numerical simulation of a jet with initial density $n_{\rm j} = 5000$ cm$^{-3}$, ten times denser than the ambient medium ($\nu = 0.1$), and with Mach number $M = 20$, corresponding to an initial jet velocity $v_{\rm j} \approx 950$ km/s. We found that this model is thermally unstable since the size of the X-ray emitting region at the head of the jet is larger than the corresponding characteristic Field length. As a consequence, the X-ray luminosity initially at values $\approx3\times10^{29}$ erg/s drops of over $2$ orders of magnitude in about $10$ years.

Higher values of the X-ray emission could be obtained in cases with: 1) higher initial jet density, $n_{\rm j}$; 2) higher ambient-to-jet density contrast, $\nu$; 3) higher initial jet velocity, $v_{\rm j}$.

To account for the first option (higher $n_{\rm j}$), we performed the same numerical simulation discussed above, but with initial jet density $10$ times greater, namely $n_{\rm j} = 5\times10^{4}$ cm$^{-3}$. In this case, we derived shock front velocity $\approx800$ km/s and X-ray luminosity ranging between $3$ and $30\times10^{29}$ erg/s and emission consistent with HH observation in general, but too high to reproduce the observations of HH 154. Such initial jet density adopted turns out to be much higher than those derived from observations of HH objects. \citet{pbn06} studied several HH objects and derived principal physical properties as their density. In particular they found that the density ranges between $400$ and $1000$ cm$^{-3}$ (see also \citealt{fl98} for density values in HH 154).

As it concerns the second and third options discussed above, based on our exploration of the parameter space Fig. \ref{fig:griglia}, we expect that for fixed initial jet velocity, the X-ray luminosity increases with increasing $\nu$ and vice versa.
Again we conclude that the jet must be less dense than the ambient medium and/or with initial jet velocity higher than $950$ km/s.
Once again our analysis leads to the conclusion that X-ray emission originating from protostellar jets (in particular from HH 154) is better reproduced by light jets with initial jet density $n_{\rm j} = 500$ cm$^{-3}$.

\section{Discussion and conclusions}
\label{Discussion and conclusions}

We presented an hydrodynamic model which describes the interaction between
a supersonic protostellar jet and a homogeneous ambient medium. The aim
is to derive the physical parameters of a protostellar jet which can
give rise to X-ray emission consistent with the recent observations of
HH objects.

In a previous paper (\citealt{bop04}), we have shown the feasibility
of the physical principle on which our model is based: a supersonic
protostellar jet leads to X-ray emission from the shock, formed at
the interaction front with the surrounding gas, consistent with the
observations in the particular case of HH\,154, the nearest and best
studied X-ray emitting protostellar jet. Here we have performed an
extensive exploration of a wide space of those parameters which
mainly describe the interaction of the jet/ambient system: the jet
Mach number, $M$, and the ambient-to-jet density contrast, $\nu$
(Fig.~\ref{fig:griglia}). These results, therefore, provide insight in
a wider set of phenomena, allowing us to study and diagnose physical
properties of protostellar jets in general, other than that in L$1551$.

One of the main results of our analysis is that only a narrow range of
parameters can reproduce observations. 
The extensive exploration of the parameter space here discussed improves and extends the previous work by allowing us to constrain the main protostellar jets parameters in order to obtain X-ray emission, best fit temperature and shock front speed consistent with experimental data.

The range of parameters which more significantly influences the jet/ambient evolution are shown in Tab. \ref{tab:range}. From a comparison with observed quantities (also shown in Tab. \ref{tab:range}, according to \citealt{br02}) it can be deduced that the parameters used in our model are consistent with observed values. Note that the values of initial jet velocity are higher than the observed values. This apparent discrepancy is due to the fact that observers measure the velocity of the knots which have already been slowed down by the interaction with the ambient medium. So we need a higher initial jet velocity to account for these lower speed values at the working surface. However in the best light jet case, we derive a kinetic power $L_{\rm mech}\approx0.3$ $L_{\odot}$, more than 2 orders of magnitude lower than the observed bolometric luminosity of HH $154$, $L_{bol}\approx40$ $L_{\odot}$ (see Tab. \ref{tab:obs}). So the jet velocity values used in our simulations lead to reasonable kinetic power. Furthermore comparing the X-ray luminosity derived in the best light jet model here discussed, we deduce that only a small fraction of the kinetic power is converted in X-ray emission: $L_{X}/L_{\rm mech}\approx (8\times10^{-5}$ $L_{\odot})/(0.3$ $L_{\odot}) = 3\times10^{-4}$.

In the following we summarize
our findings:

\begin{itemize}
\item {\em Light jet} ($\bf\nu > 1$).\\ The light jet cases which
  reproduce the X-ray emission and optical proper motion observed are
  those with initial Mach number $M\ge300$ and ambient-to-jet density
  contrast $10 \leq \nu < 100$.  For Mach numbers lower than $M = 300$,
  the X-ray luminosity derived from our simulations is lower than the
  minimum value observed in protostellar jets ($(L_{\rm X})_{\rm obs} >
  10^{28}$ erg s$^{-1}$) and, in some cases, with a transient behaviour
  due to thermal instabilities. The values $M = 300$ and $\nu = 10$
  provide the best case which reproduces the HH\,154 observations in
  terms of best fit temperature, emission measure and X-ray luminosity. We
  also predict a substantial proper motion.

\item {\em Equal-density jet} ($\bf\nu = 1$).\\ The three equal-density
  jet cases analyzed allow us to constrain the initial Mach number of
  an equal-density protostellar jet to reproduce observations: $30 <
  M < 300$. From the equal-density model with $M = 100$ we derive
  shock front velocity and X-ray luminosity consistent with HH objects
  observations in general (see Tab.~\ref{tab:obs}).

\item {\em Heavy jet} ($\bf\nu < 1$).\\ To reproduce some characteristics
  of the observations, in the heavy jet scenario, we need initial jet Mach
  number $M\approx30$ and initial jet density much higher than the
  ambient density, $\nu \leq 0.03$, i.e. a jet $30$ times denser than
  the medium or more. Lower initial Mach number in some cases leads
  to thermal instability which suppress X-ray emission about $5$ years
  since the beginning of the interaction between the protostellar jet and
  the ambient medium. A jet $30$ times denser than the ambient medium
  ($\nu = 0.03$) and with initial Mach number $M = 30$ predicts emission
  from a million degrees plasma. Although its $v_{\rm sh}$ is too high
  and its $L_{\rm X}$ is too low with respect to those observed in
  HH objects in general (see Tab.~\ref{tab:obs}) and in HH\,154 in
  particular (\citealt{ffm02}), we cannot reject the possibility of new
  more sensitive observations which may show fainter emission not yet
  detected in X-ray emitting HH objects.
\end{itemize}

Here we discussed the best-fit models of light, heavy and equal-density
jets in best agreement with experimental results from HH objects in
general: a light jet with $M = 300$ and $\nu = 10$; an equal-density
jet with $M = 100$ and $\nu = 1$; and a heavy jet with $M = 30$ and $\nu
= 0.03$.

For each case, we analyzed the evolution of the mass density and
temperature spatial distributions derived from our model, the shock
front proper motion and its spectral properties, the X-ray emission
and its stability.  In each best-fit model the interaction between the
supersonic protostellar jet and the unperturbed ambient medium leads to
the formation of a hot and dense cocoon surrounding the jet and smoothed
by the thermal conduction. Just behind the shock front, there is a hot
and dense blob from which the harder and bright X-ray emission originates:
the strong interstellar absorption suppresses the softer component due to
the cocoon. In all cases examined, the X-ray emitting region is thermally
stable, i.e. thermal conduction effects prevent the collapse of the
source due to radiative cooling, and shows a detectable proper motion.

To compare our findings with HH\,154 observations, we have rejected any
equal-density and heavy jet case which show too high shock front velocity
($v_{\rm sh}\approx600$ km s$^{-1}$ and $\approx1900$ km s$^{-1}$,
respectively) and too low X-ray luminosity ($L_{\rm X} \approx 10^{29}$
erg s$^{-1}$ and $\approx 10^{28}$ erg s$^{-1}$) with respect to $v_{\rm
sh}$ and $L_{\rm X}$ values observed in HH\,154 ($v_{\rm sh}\approx 500$
km s$^{-1}$ and $L_{\rm X} = 3\times10^{29}$ erg s$^{-1}$).

In the best light jet case, we derived a particle density $n\approx6500$ cm$^{-3}$ and a velocity $v_{sh}\approx500$ km s$^{-1}$ for the X-ray emitting region, leading to a momentum $mv = 2\times10^{-6}$ $M_{\odot}$ km s$^{-1}$. This value is consistent with the upper limit $7\times10^{-4}$ $M_{\odot}$ km s$^{-1}$ obtained for HH $154$ (\citealt{fl98}).
This leads to the conclusion that the protostellar jet cannot drive the molecular outflow, whose momentum has been estimated to be between 0.15 and 1.5 $M_{\odot}$ km s$^{-1}$ (\citealt{fl98}). The relations for the mass loss rate and the mechanical luminosity

$$\dot{M} = 1.8\times10^{-9} M_{\odot}/yr \left(\frac{r_{\rm j}}{5\times10^{14} cm}\right)^{2} $$
\begin{equation}
\left(\frac{n_{\rm j}}{500 cm^{-3}}\right) \left(\frac{v_{\rm j}}{1400 km/s}\right)
\label{eq:mass-loss}
\end{equation}

\begin{equation}
L_{\rm mech} = 0.3 L_{\odot}\left(\frac{r_{\rm j}}{5\times10^{14} cm}\right)^{2} \left(\frac{n_{\rm j}}{500 cm^{-3}}\right) \left(\frac{v_{\rm j}}{1400 km/s}\right)^{3}
\label{eq:mass-loss}
\end{equation}
\noindent
lead, respectively, to values 3 and 2 orders of magnitude lower than expected in CO outflows (\citealt{cb92}). This result supports the conclusion discussed by \citet{fl98} that the jet origin is probably different from that of the CO outflow.

We estimated the values of the momentum, mass loss rate and mechanical luminosity for each model which better reproduce the X-ray observation of HH objects shown in Tab. \ref{tab_mod}. We derived: $\dot{M}\sim10^{-9}$ $M_{\odot}$/yr; $L_{\rm mech} = (0.3 - 1.7)$ $L_{\odot}$; $mv \la 10^{-6}$ $M_{\odot}$ km s$^{-1}$. These values are several orders of magnitude lower than those observed in CO outflows (in HH 2, HH 154 and HH 80/81, \citealt{mcn99}; \citealt{fl98}; \citealt{ysk89}). 
From the comparison between our results and the observations of CO outflows, we conclude that the simulated protostellar jets which best reproduce X-ray observations cannot drive molecular outflow.

On the base of our analysis, we conclude that the light jet scenario,
$10$ times less dense than the ambient medium ($\nu = 10$) and with
initial Mach number $M = 300$, with $v_{\rm sh}\approx 500$ km s$^{-1}$
and $L_{\rm X} \approx 3\times10^{29}$ erg s$^{-1}$, is the best case
to reproduce the HH\,154 observations.  This is also supported by the
optical observations of HH\,154 (\citealt{fld05}) from which the light
jet scenario can be deduced, according to the \citet{har89} model.

\begin{figure}[!t]
\centerline{\psfig{figure=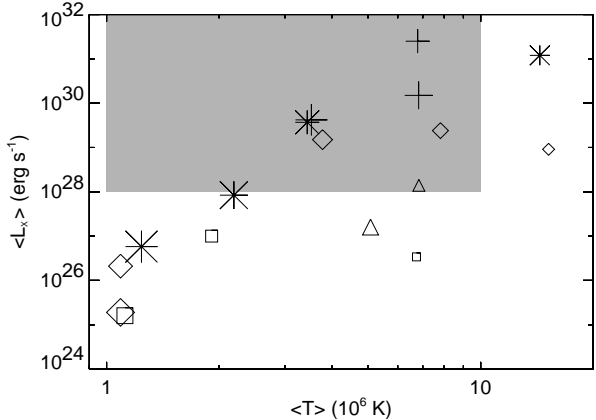,width=8cm}}
\caption{Mean value of the X-ray luminosity as a function of the mean
value of the best fit temperature, derived from our model synthesized
spectra. Crosses correspond to Mach number $M = 1000$, stars to $M =
300$, diamonds to $M = 100$, triangles to $M = 30$ and squares to $M =
10$. Increasing size of each symbol corresponds to increasing value of the
ambient-to-jet density contrast, $\nu = 0.01$-$300$. The shaded region
refers to the ($L_{\rm X}$, $T$) values consistent with observations of
HH objects in general.}
\label{fig:L-T}
\end{figure}

More in general, in Fig.~\ref{fig:L-T} we show the values of $L_{\rm X}$
vs. $T$ derived from the spectra synthesized from our model as a function
of different values of $M$ and $\nu$. Crosses mark cases with $M = 1000$,
stars with $M = 300$, diamonds with $M = 100$, triangles with $M = 30$
and squares with $M = 10$. Bigger symbol sizes correspond to higher
values of the ambient-to-jet density contrast, $\nu$, in the range 0.01
to 300, as in Fig.~\ref{fig:griglia}. The shaded zone marks the range
of parameters consistent with observation (Tab.~\ref{tab:obs}). We
have chosen $L_{\rm X}$ one order of magnitude lower than the minimum
observed so far to account for fainter sources.

For a fixed value of $M$, $L_{\rm X}$ and $T$ decrease with increasing
$\nu$ and, for a fixed value of $\nu$, $L_{\rm X}$ and $T$ increase
with increasing $M$. From the figure it is possible to derive $M$,
$\nu$ (velocity and density) of the protostellar jet to be compared
with observations (in terms of $L_{\rm X}$ and best fit $T$).
Predictions about fainter not yet discovered sources can also be made.

Furthermore the results derived from the variation of the initial jet density parameter, $n_{\rm j}$, discussed in Sect. \ref{Varying the jet density parameter}, lead to the conclusion that, even with different initial density values, a protostellar jet must be less dense that the ambient medium and with high initial velocity ($1000$ km/s or more) in order to reproduce HH objects observations.

Our model predicts in all cases a significant proper motion of the
X-ray source, with values which, in the case of HH\,154 would be
measurable with \emph{Chandra}, providing a clear test of the model
scenario. 

As discussed by \cite{fbm06}, a 100 ks observation was
performed in 2005, showing, when compared with the 2001 observation, a
more complex scenario, i.e.\ both a moving and a stationary source
were detected in HH\,154, giving the source, in 2005, a ``knotty''
appearance. Thus, while a traveling shock (likely based on the basic
physics explored in the present work) is apparently present in
HH\,154, the source structure is more complex. 

The comparison between our model of a continuous supersonic jet through an unperturbed surrounding medium and the new \emph{Chandra} data, discussed in \citet{fbm06}, shows that the model reproduces most of the physical properties observed in the X-ray emission of the protostellar jet (temperature, emission measure, etc.). At the same time, it fails to explain the complex evolving observed morphology, showing, most likely, that the jet is not continuous.

A possible scenario,
which we will test in the future, is based on similar physics, but in
the presence of pulsating jets (instead of the continuous jet here
examined), or alternatively on interaction between the jet and an
inhomogeneous ambient medium which can lead to the knotty structure
observed inside the jet itself. We also plan to explore other physical mechanisms, different from the
moving shock at the tip of a supersonic jet, stimulated by the above
mentioned \emph{Chandra} observations of HH\,154, such as steady shocks
formed at the mouth of a de Laval nozzle. New observations of the
evolution of HH\,154 will however be necessary to understand the
phenomenon and to further constrain the model scenario.

The continuous jet model discussed here is a useful and  necessary building block toward more complex (e.g. with a discontinuous time profile) models. The X-ray emission from a pulsed jet, for example, will still take place at the shock front, and will thus be based on the same physical effects and principles as observed in a continuous jet. Our (simpler) continuous jet model is a very useful tool to infer the right parameter values to use in future, more complex models needed to also reproduce the observed morphology: using our continuous jet model results, shown in Fig. ~\ref{fig:griglia} and in Fig. ~\ref{fig:L-T}, it is possible to derive the initial jet velocity and ambient-to-jet density ratio needed for the new bullets to produce to X-ray luminosity and best fit temperature consistent with observations.

A limitation of our hydrodynamic model is the hypothesis of pressure balance between the jet and the ambient medium, in order to obtain the observed jet collimation. Most models of jet collimation suggest the presence of an organized ambient magnetic field which is known to be effective in collimating the plasma.
As a follow-up of our analysis, we are developing an MHD model of protostellar jet that will allow us to relax the assumption of an initial pressure
equilibrium. The comparison of the MHD model results with the X-ray observations will provide a fundamental tool to investigate the role of the magnetic field on the protostellar jet dynamics and emission.

\begin{acknowledgements}

We would like to thank M. Fridlund for stimulating discussions.
The software used in this work was in part developed by the DOE-supported
ASCI/Alliances Center for Astrophysical Thermonuclear Flashes at the
University of Chicago, using modules for thermal conduction and optically
thin radiation constructed at the Osservatorio Astronomico di Palermo. The
calculations were performed on the cluster at the SCAN (Sistema di
Calcolo per l'Astrofisica Numerica) facility of the INAF -- Osservatorio
Astronomico di Palermo and at CINECA (Bologna, Italy).
This work was partially supported by grants from CORI 2005, by Ministero
Istruzione Universit\`{a} e Ricerca and by INAF.

\end{acknowledgements}

\bibliographystyle{aa}
\bibliography{references}

\newpage
\appendix

\section{Synthesizing the X-ray spectra}
\label{synthesizing the X-ray spectra}

From our 2-D numerical simulations, we synthesized the absorbed focal plane spectra to be compared with observations by using the following procedure.

As a first step, from the integration of the hydrodynamic equations ~\ref{eq:massa-1}, ~\ref{eq:momento-1} and ~\ref{eq:en+r+c-1}, we derive the temperature and density 2-D distributions in the computational domain.

We reconstruct the 3-D spatial distribution of these physical quantities by rotating the 2-D slabs around the symmetry axis. Then we derive the emission measure, defined as $EM = \int n_{e} n_{\rm H} dV$
(where $n_{e}$ and $n_{\rm H}$ are the electron and hydrogen densities,
respectively, and $V$ is the volume of emitting plasma).

From the 3-D spatial distributions of $T$ and $EM$, we derive
the distribution of emission measure $EM(T)$ for the computational
domain as a whole or for part of it: we consider the temperature range
$[10^{3}-10^{8}]$ K, divided into $74$ bins equispaced in $\log T$;
the total $EM$ in each temperature bin is obtained summing the emission measure
of all the fluid elements corresponding to the same bin.

From the $EM(T)$, using the MEKAL spectral code (\citealt{mgv85}) for optically thin plasma, we derive the number of photons in the i-th energy bin as

\begin{equation}
I_{i} = \frac{1}{4\pi D^{2}}\sum_{k}\int_{E_{i}}^{E_{i+1}}\frac{P(T_{k}, E) EM(T_{k})}{E} dE
\end{equation}
\noindent
where $D$ is the distance of the object, $E_{i}$ is the energy in the i-th bin, $P(T_{k}, E)$ describes the radiative losses as a function of the energy and of the temperature in the k-th bin.

To compare our model results with observations, we synthesize the focal plane spectrum, $C_{i}$, as predicted to be observed with the \emph{Chandra}/ACIS-I or
XMM-\emph{Newton}/EPIC-pn X-ray imaging spectrometers taking into account the spectral instrumental response:

\begin{equation}
C_{i} = \frac{t_{exp}}{4\pi D^{2}} \sum_{k} \int_{E_{i}}^{E_{i+1}}A(E)M(i,E)\frac{P(T_{k},E)EM(T_{k})}{E}dE
\end{equation}
\noindent
where $t_{exp}$ is the exposure time, $A(E)$ is the effective area and $M(i,E)$ is the instrumental response.

Finally we take into account the interstellar medium absorption column density, $N_{\rm H}$ (\citealt{mm83}), and we analyze the absorbed focal plane spectrum with XSPEC V11.2 in order to compare our findings with published experimental results.

\end{document}